\documentclass[amssymb,prb,preprint,aps]{article}
\usepackage{epsfig}
\usepackage{amsmath,amssymb,enumerate,mathrsfs
}
\usepackage {latexsym}
\usepackage{amssymb}
\usepackage{graphics}

\begin{document}
\title{The Nature of the magnetism-promoting hole state in the
prototype magnetic semiconductor GaAs: Mn}
\author{V. Fleurov , K. Kikoin \\
Raymond and Beverly Sackler Faculty of Exact Sciences, \\ School of
Physics and Astronomy, \\ Tel-Aviv University, Tel-Aviv 69978
Israel. \and A. Zunger\\ University of Colorado, Boulder, Colorado
80309, USA} \maketitle
\begin{abstract}
Recent experiments \cite{[1]} suggest that the ferromagnetism (FM)
in GaAs: Mn is determined by the impurity band rather than holes in
the valence band. We discuss here the physical mechanism of FM
mediated by the carriers in impurity band, where the Mn d-level play
a crucial role. The theory is based on the first principle approach.

\end{abstract}

\maketitle

The paradigm system that combines ferromagnetism (FM) with
semiconductivity involves Mn$^{(2+)}$ impurity ions substitution for
Ga$^{(3+)}$ atoms  in GaAs \cite{[2],[3],[4],[5],[6],[7]}. Such
acceptor substitution creates a hole that interacts with the local
moment of d$^5$ Mn. This doping-induced magnetism could lead to
electrical control of FM, to the potential benefit of
spin-electronics (spintronics). The nature of the ferromagnetism,
including its dependence on the hole concentration and on that of
the Mn ions depends, however, on the physical nature of the hole
state. One view -the "host like hole" model \cite{[2],[3],[4],[5]}
has been that the hole resides inside the GaAs valence band. Such
view would permit the use of the language of GaAs semiconductor
physics (s-p bonding, extended wave functions, RKKY exchange) in
analyzing the ensuing magnetism and its dependence on concentration
of the relevant species. This scenario, underlying most Model
Hamiltonian treatments of the problem \cite{[3],[5]}, was inspired
by the previously known case of isovalent  Mn doping of CdTe, where,
on account of the host metal atom Cd$^{(2+)}$ having the same charge
as the magnetic impurity ion Mn$^{(2+)}$, hole formation required
additional doping. Such doping was accomplished by conventional
hydrogen-like dopants (extended wave function in the effective mass
approximation), leading to the expected host-like hole behavior
underlying delocalized, effective-mass dopants. The different,
"Impurity Band" view \cite{[6],[7]} emerged from the assumption that
Mn doping is unlikely to be hydrogen-like, as it introduces into
GaAs a fundamentally new (d) orbital type, absent from the (s ,p)
host. Then it is not obvious {\em a priori}, whether the hole will
carry the identity of the host or that of the impurity; electronic
structure calculations were needed to make this judgment. First
principles calculations \cite{[6],[7]} have suggested that the hole
resides in an impurity band above the host valence band. This view
implies that the magnetism could not be described in the language of
host semiconductor physics alone, but rather by that related to the
localized d band of Mn, hybridized with $t_2$ states of the host.
Understanding which of the two views of the nature of the hole state
is correct is important for deciding the pertinent guidelines for
optimization of spintronic devices fabricated from FM
semiconductors.

Some experimental observables related to the GaAs:Mn system  are not
very sensitive to the nature of the hole state, and could be
explained either way. Examples of such non-crucial experiments
include effects reflecting predominantly the existence of local
moments of Mn interacting with some background carriers in the
Kohn-Luttinger s-p bands, including magneto-transport,
magneto-optics, thermoelectrical effects and other phenomena related
to itinerant rather than to localized carriers near the top of the
valence band. Remarkably, however, a recent crucial experiment [1]
seems to have settled this debate in favor of the Impurity Band view
on the mechanism of FM ordering by measuring independently the net
densities of holes and that of the Mn ions, showing that the Fermi
level resides above the valence band, inside the impurity band and
that the Curie temperature $T_c$ is controlled by this position
rather than by the density of nearly free carriers as in the
host-like-hole view. A Cover Story \cite{[8]} echoed this view.
While explaining that "the compass is pointing in the direction of
impurity band scenario", this piece \cite{[8]} expressed the concern
that some experimental features demonstrated by the most metallic
samples are unclear as to their compliance with a particular hole
model. We point out here that there are fundamental
model-independent reasons for assertion that the placement of the
hole in an impurity band (above the host valence band) holds both in
the Mn- dilute  limit (on the insulating side) and in the high
concentration limit $n_{Mn} > 0.1$ (on the metallic side).

\begin{description}

\item{(i)} The Mn-induced acceptor level in III-V semiconductors is a
result of orbital hybridization leading to a deep acceptor-like
impurity band, not effective-mass like:  Studies of the chemical
trends in transition metal (TM) doped III-V semiconductors in a
dilute doping limit \cite{[9],[10]} have shown that these trends are
determined by the strong d-p hybridization between the d-orbitals of
TM ion and the p-orbitals of its nearest anion neighbors. The
ensuing states are determined by the relative location of the atomic
3d levels of TM ions and the center of gravity of the heavy hole
valence band. This results in two types impurity-related states: the
bonding, TM localized "Crystal Field Resonance" (CFR) and the
antibonding state called "Dangling Bond Hybrid" (DBH), in which the
TM d state hybridizes with the vacancy-like dangling bonds. (See
figure \ref{1}.)
\begin{figure}[ht]
\includegraphics[width=8cm]{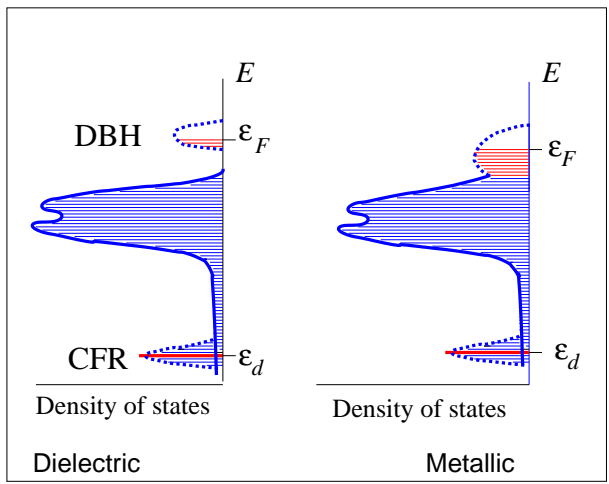}
\caption{Density of electron states in dielectric (left) and
metallic (right) phases of Ga(Mn)As. Two impurity-related peaks in
DOS form a covalent pair CFR (related to the Mn($d^5/d^4$) level) -
DBH (impurity band above the top of the valence band. Only the t2
partial contribution in the heavy hole band DOS is schematically
shown.The CFR/DBH structure originated from the local dp bonding
survives at any level of doping} \label{1}
\end{figure}
The ferromagnetic spin arrangement leads to energy lowering (ground
state) since more spins are occupying the bonding than the anti
bonding state \cite{[6],[11]}. In a given host semiconductor, the
energetic positions of CFR and DBH vary in a universal way with the
atomic number of the TM \cite{[6],[9],[10],[11]}. On the other hand,
considering different III-V semiconductors such as GaN-GaP-GaAs-GaSb
with their valence band maximum (VBM) aligned according to their
band offsets, the position of the DBH-CFR levels is approximately
constant, as shown in references \cite{[6],[12]}. These chemical
trends unequivocally point to the d-p hybridization, rather than
hydrogen-like acceptors, as the principal mechanism of formation of
acceptor levels and impurity bands. Critically, Mn in GaAs, creates
a DBH state outside the host valence band (inside the gap); the fact
that this level is only $~$ 0.1 eV above the VBM should not be taken
to imply that this is a shallow, host-like level, since its wave
function is indeed composed from a multitude of {\bf k}-points
(unlike effective-mass or ${\bf k}\cdot{\bf p}$ description) and
contains d-character absent from the host crystal. In the more
extreme case of Mn in GaN, where the CFR acceptor transition is very
deep in the gap (~ 1.4 eV above the VBM) and cannot possibly lead to
ionizeable free holes, the magnetism clearly cannot be described by
the host like hole s-d-exchange theory \cite{[13]} with inter-site
spin exchange. (See also discussion in \cite{[16]}).

\item{(ii)} The model of impurity potential used in the k?p approach to
deduce a delocalized nature of the Mn states is not appropriate for
the 3d impurities: Ref. [13] attempted to describe Mn in GaAs as a
square well potential interacting with the host bands; in the
presence of many such wells the bound state) spreads out, creating
the impression of a host-like resonance impurity state. This
approach relies also on the RKKY mechanism to account for the FM,
which was shown to be inadequate \cite{[14],[15]}. The description
of resonance impurity scattering by means of potential scattering
[13] is misleading because it lacks the d-orbital nature of the real
Mn state with its specific Mn d orbital energy. In turn, the d-p
hybridization mechanism [(i) above] with its correct Mn d orbital
energy places the state outside the host bands and  is robust, in
the sense that it  cannot be significantly  modified by any kind of
screening, or disorder effects.  In fact, disorder cannot destroy
the local chemical bonds. The immediate consequence of this fact is
that even for concentrated Mn, on the metallic side, where the
impurity band is merged with the valence band and there is no gap
for charge transport, the top of this impurity band is still formed
by strongly hybridized d-p-orbitals (see Fig. \ref{1}).

\item{(iii)} The impurity band picture correctly describes the dependence of
$T_c$ on the Fermi energy position: In accordance with above picture
the Fermi level is pinned in the impurity band region of strong d-p
hybridization both in insulating and metallic states. This impurity
band modifies the host crystal density of states (DOS) generating a
Lorentzian-like impurity band above the host valence band maximum.
This DOS is responsible for the details of the hole-mediated Zener
superexchange as shown in Refs. \cite{[16],[17]}.

(iv) In GaAs:Mn films the Fermi level depends on the ratio between
substitution and interstitial Mn-related defects, and the optimum
concentration corresponds to half-filling of the impurity band.
Thus, the dependence of $T_c$ on the effective carrier concentration
x has the 'dome-shaped' function observed in experiment (Ref.
\cite{[1]}, Fig. 1). This type of dependence follows from the model
calculations \cite{[16],[17]} and first-principles supercell
calculations \cite{[9]} based on the mechanism of d-p-hybridized
CFR-DBH states, and therefore on the pinning of these states to the
universal energy scale (the valence band offset). A dome-like
dependence $T_c(x)$ obtained within the model of impurity
ferromagnetism due to Zener-like double exchange via the impurity
band \cite{[16]} can be seen in Fig. 2 of that paper.

\item{(v)} Two recent experimental findings unambiguously support the
statement that the d-p-hybridization is responsible not only for the
Zener exchange but also for the shape of the DOS near the top of the
valence band in metallic ferromagnetic GaMnAs. These are the
dome-shaped Tc(x) (Ref. \cite{[1]} Fig. 1), and the absence of Drude
peak in the infrared conductivity of "metallic" samples
(\cite{[18]}, Fig. 10), (\cite{[19]}, Figs. 1 and 3), which
indicates the absence of free charge carriers at the Fermi level in
the samples.

\item{(vi)} The mechanism of d-p hybridized hole states above the top
of the valence band provides guideline for optimization of devices
fabricated from FM semiconductors: There are some important
practical consequences of the change of FM model from host-like-hole
to impurity band hole: (a) one should look for the fabrication and
annealing regimes which favor the optimal half-filling of Mn-related
impurity band; (b) the narrower is the band, the higher Tc is
expected. To make this band narrower, the fabrication of
heterostructures with spatially quantized impurity bands may be
useful.

\end{description}

To conclude, we believe that the recent experimental findings
together with the fundamental quantum-mechanical properties of Mn in
III-V semiconductor hosts unambiguously point out on the localized
nature of holes mediating the Zener exchange in GaMnAs .

\end{document}